\documentclass{iau} 
\usepackage{graphicx}
\usepackage{natbib}
\usepackage{aas_macros}

\title[Photoionization models of PNe] 
{Photoionization models of Planetary Nebulae}

\author[Christophe Morisset]   
{Christophe Morisset}

\affiliation{
Instituto de Astronomia \\ 
Universidad Nacional Autonoma de Mexico \\
Circuito Exterior, area de la Investigacion Cientifica\\
Ciudad Universitaria, C.P. 04510, CDMX, Mexico \\ 
email: {\tt chris.morisset@gmail.com} 
}

\pubyear{2017}
\volume{323}  
\jname{Planetary Nebulae: Multi-wavelength probes of stellar and galactic evolution}
\editors{Xiaowei Liu, Letizia Stanghellini, and Amanda Karakas.}

\begin{document}

\newcommand{\ion}[2]{#1~{\sc #2}}
\newcommand{\hi}{\ion{H}{i}}
\newcommand{\hh}{\ion{H}{ii}}
\newcommand{\hei}{\ion{He}{i}}
\newcommand{\heii}{\ion{He}{ii}}
\newcommand{\cii}{\ion{C}{ii}}
\newcommand{\nn}{\ion{N}{i}}
\newcommand{\nii}{\ion{N}{ii}}
\newcommand{\neii}{\ion{Ne}{ii}}
\newcommand{\ariii}{\ion{Ar}{iii}}
\newcommand{\oi}{\ion{O}{i}}
\newcommand{\oii}{\ion{O}{ii}}
\newcommand{\oiii}{\ion{O}{iii}}

\maketitle

\begin{abstract}
The understanding of astronomical nebulae is based on observational data (images, spectra, 3D data-cubes) and theoretical models. In this review, I present my very biased view on photoionization modeling of planetary nebulae, focusing on 1D multi-component models, on 3D models and on big database of models.

\keywords{methods: numerical, planetary nebulae: general, ISM: abundances}
\end{abstract}

\firstsection 
\section{Introduction}
This talk could well be subtitled "Fifty Years of Nebular Modeling" inspired by the introduction of the review by \citet{1989Harrington_131} at the IAU meeting on Planetary Nebulae (PNe) in Mexico in 1987. The first published models of PNe are indeed almost 50 years old: \citet[][on theoretical study on the ionization structure of nebulae]{1967Goodson_zap66}, \citet[][making models for IC418 and NGC7662]{1968Flower_aple2}, \citet[][on ionization structure and thermal stability]{1968Harrington_apj152} and \citet[][presenting models at the IAU PNe symposium 34]{1968Williams_34}. Impressively, most of these results were obtained using atmosphere models  \citep[from][]{1965Bohm_zap61,1966Bohm_zap63} instead of simple Planck functions for the ionizing spectral energy distributions (SEDs).

Planetary nebulae (as well as any other component of the ionized interstellar medium like \hh\ regions, nebulae around AGNs, WR bubbles, novae and supernovae remnants) can be considered as active filters to the ionizing radiation: they transform the almost continuous incident stellar Lyman continuum radiation into a spread-out continuum spiked with emission lines produced following recombinations and collisional excitations \citep[more on theory in][]{2006Osterbrock_}.
Photoionization codes solve the equations of ionization and thermal equilibria for gas and dust particles and compute the radiation transfer. This allows to the determination of the ionic fractions, the electron temperature and the line and continuum emissivities in each cell of the model. One can see a photoionization code as a detailed quantitative summary of our knowledge of gas-photon interactions in the interstellar medium.

Today's most commonly used photoionization codes are: Cloudy \citep[][and the recent conference to honor Gary Ferland, Mexico, 2016]{2013Ferland_rmxa49}, MAPPINGs IV \citep[][this code also deals with shocks]{2013Dopita_apjs208} and MOCASSIN \citep[][this code being Monte Carlo full 3D]{2003Ercolano_mnra340, 2008Ercolano_apjs175}. The python library pyCloudy \citep{2013Morisset_, 2014Morisset_} allows to manage the Cloudy code and to run pseudo-3D models \citep[][and section \ref{sec:3d-modeling}]{2005Morisset_mnra360}. Notice the impressive effort by Gary Ferland and collaborators to provide a very complete and useful manual to the use of Cloudy: this is certainly one of the most important ingredients to the success of this code (among the quality of the code itself, of course).

\citet[][at the IAU PNe symposium 180]{1997Pequignot_180a} centered his review on the need for ultra deep spectra to identify and measure lines from the third row elements. This is still a challenge and recent efforts have lead to observations (and sometimes first identifications) of emission lines from Se, Kr, Rb, Xe, Ca, K, Cr, Mn, Fe, Co, Ni, and Cu \citep{2015Garcia-Rojas_mnra452}, and Cd, Ge and Rb in the infra-red \citep{2016Sterling_apjl819}.
\citet[][at the IAU PNe symposium 234]{2006Ercolano_234} focused her review on advances obtained in 3D modeling, models with dust and PDR regions. This is still of actuality, see the following sections. Both point out the importance of atomic data in obtaining trustable models.

In the following pages I will first describe the way photoionization codes are working, and what they are useful for. I will then focus on a few illustrations of tailored models, big database of models and 3D models.

\section{Photoionization models : ingredients and outputs}

To run a photoionization model, one needs to give a description of the nebula to be modeled in terms of the ionizing radiation field and of the properties of the ionized gas. The first item involves the intensity and the shape of the ionizing stellar continuum. It can be a simple black body or a more complex result from a stellar atmosphere modeling code, for example the models by \citet{1997Rauch_aap320}. The shape is then determined by the effective temperature, the gravity, and the stellar metallicity). On the side of the ionized gas, one must specify: 1) the density, usually in form of the hydrogen density, 2) the elemental abundances, 3) the dust/gas ratio for each type of dust -- all these parameters may vary with the position in the nebula --, and 4) the total mass of the nebular gas, which leads to matter- or radiation-bounded nebulae. Distance to the nebula may be needed if one wants to compare the model with absolute fluxes. A velocity field must be provided to allow comparison with emission line profiles or Position-Velocity diagrams.

To this specific description of the object one must add a set of atomic data to describe quantitatively every process that occurs in the nebula. This includes photoionization cross sections, recombination coefficients including dielectronic processes, transition probabilities, collision strengths, charge exchange coefficients, molecular formation and dissociation rates as well as dust optical properties. In the case of Cloudy, the data directory weights 432 Mb, while the source directory is "only" 8.2 Mb size (this still represents almost 200,000 lines of C++ code...). A great effort was made by Cloudy's team to define a new (universal?) format for atomic data, as described by \citet{2015Lykins_apj807}. The models can be used to determine or constrain atomic data, as done for the charge transfer rates in a pioneering work by \citet{1978Pequignot_aap63}. 
More recently \citet{2012Escalante_mnra426} determine the contribution of the fluorescence to the IC418 recombination spectra of \cii, \nn, \nii, \oi, and \oii. 

The results of a photoionization code applied to a PNe are (among others): the electron temperature, the ionic fractions, the continuum and line emissivities, and the dust grains temperatures. For all these parameters, the values are determined for each cell in the nebula\footnote{A cell can be a spherical shell between R and R + $\Delta$R in the case of a spherical 1D model, or an element of a Cartesian grid in the case of a 3D model.}. It is important to realize that the electron temperature is a result of the code, not an input. If the line ratios of the temperature diagnostics are not reproduced, the user must change something in the model: this may be the SED hardness that modifies the heating, or the abundances that act on the cooling, or consider additional process that modifies the heating or the cooling. Otherwise, the determination of the abundances from the model is meaningless.

\subsection{Limitations of static photoionization models}

The limitation of the photoionization models are basically the ones of their ingredients, for example: 1) the input SEDs are obtained from atmosphere models that have never been confronted to reality in the -- unobservable -- ionizing domain, 2) the atomic data are still incomplete and suffer inconsistencies between different authors, 3) the presence of additional physical process that are not always taken into account ($\kappa$-distribution, magnetic reconnection, shocks) that may explain some discrepancies.

\section{Classical 1D models}

\subsection{Tailored models}

They are used to determine the properties of a given object, e.g abundances. A converged model is supposed to reproduce all available observations. 

As an example, one can consider the detailed model of IC418 by \citet{2009Morisset_aap507}. In this work, we combined a tailored stellar model with a tailored nebular model. This allowed us to determine the distance to the object, using its age determined from its size and expansion velocity and the stellar age, from stellar evolutionary tracks and stellar properties. Without these information, one cannot resolve the degeneracy between the distance and filling factor \citep[see][for details]{2009Morisset_aap507}. 
The nebula is the only witness of the ionizing SED. In this work, we determined the importance of a correct description for h$\nu > 40$~eV ([\ariii] and [\neii] lines probe it).

Another example is the detailed model of TS01 by \citet{2010Stasinska_aap511}. Without any observational constraint on the electron temperature (i.e. no [\oiii]4363 nor [\nii]5755 observed intensity), it has still been possible to obtain a rather acceptable determination of the abundances: this is possible only in the case of very strong constraints on almost every other parameter: very good stellar observations, nebular images and spectra (IR, optical and UV). And the use of a 3D model to take into account the morphology and the slit positions (see below).

Some models to study the properties of dust from its infrared (IR) emission are shown by e.g. \citet{2015Otsuka_apjs217}. A detailed model must include the dust emission coming from the PDR (if existing), which can be constrained by [\oi] and [\cii] IR lines.



\subsection{Grid of models}

They are are based on toy models and are used to study some specific effects or to determine some properties when varying parameters. 

The most recent and complete grid of PNe models is the one included in the Mexican Million Models database \citep[3MdB][]{2015Morisset_rmxa51} under the references 'PNe\_2014', 'PNe\_2014\_c13' (subset of the previous runs with Cloudy version 13.03) and 'PNe\_2016' (refined mesh in effective temperature), with a total of 625,974 models, in which were varied the effective temperature, the stellar luminosity, the nebular inner radius, the gas density, the O/H abundance, the density law, the nebular mass and the presence or not of dust. An easily accessible criterion allows the user to select the most realistic models (92,513 models). The database was used to calculate PN ionization correction factors (ICFs) by \citet{2014Delgado-Inglada_mnra440} -- see below. It can also be used to look for models that fit some line ratio and thus determine the PN properties, like the stellar effective temperature, as done by \citet{2016Garcia-Rojas_aap586}.

\subsection{Abundance determinations}

The direct method is used to determine elemental abundances, in a 3 steps way: 1) determination of the physical properties: electron temperature and density (multiple zones can be considered), 2) determination of the ionic abundances, and 3) determination of the elemental abundances by applying ICF if needed.

One of the main source of uncertainties in this method is certainly linked to the ICFs. Perhaps the most cited PNe photoionization models are those used by \citet{1994Kingsburgh_mnra271} to determine famous ICFs for N, O, C, Ne, Ar, and S. The 10 PNe models their ICFs are based on should have been described in a never published paper! New ICFs have been recently published by \citet{2014Delgado-Inglada_mnra440} where the uncertainties are also specified. 

Another way to determine abundances is to compute photoionization models that fit all the observables: in this case one has to carefully define the input parameters, which is more complicated than applying the direct method. But there is no need for ICFs as the models compute the ionization state for all the elements at once. One can also try to find the set of models that fit the observables within a certain tolerance, which naturally gives us the uncertainties in parameter determination. This has been done for example by \citet{2016Esteban_mnra460} for WR bubbles. Ad-hoc ICFs have been obtained for each object, by fitting [\oiii]/[\oii], \hei/\hi\ and \heii/\hi\ line ratios with models from 3MdB (sse previous section).


\section{3D modeling}
\label{sec:3d-modeling}
Current modern observational facilities include 3D spectra (or IFUs for Integral Field Units) like VIMOS \citep{2002Le-Fevre_The-109}, 
MUSE \citep{2010Bacon_7735}, MEGARA \citep{2014Gil-de-Paz_9147}, or SITELLE \citep{2014Brousseau_9147}. The spatial mapping allows a detailed study of the morphology and emission line profiles, leading in particular to the determination of the velocity fields of each region where the emitting ion is dominant. These improvements in observational capacities require corresponding theoretical tools needed to model the nebulae in 3D.
The first 3D photoionization models are indeed from \citet{1970Kirkpatrick_apj162}, that included axially symmetric nebulae and optically thin inhomogeneities.

3D models are used for different purposes: 1) reproducing the morphology given by imaging the object, 2) taking into account the size and the position of the spectrometer aperture, 3) constraining the velocity field by high spectral resolution observations. Examples of 3D models can be found in \citet{2008Zhang_486}, \citet{2008Morisset_rmxa44}, \citet{2011Wright_mnra418}, \citet{2013Monteiro_aap560}, \citet{2013Danehkar_mnra}, and \citet{2016Akras_mnra457}.

\subsection{Aperture effects}

\begin{figure*}[ht]
\begin{center}
 \includegraphics[width=4.5in]{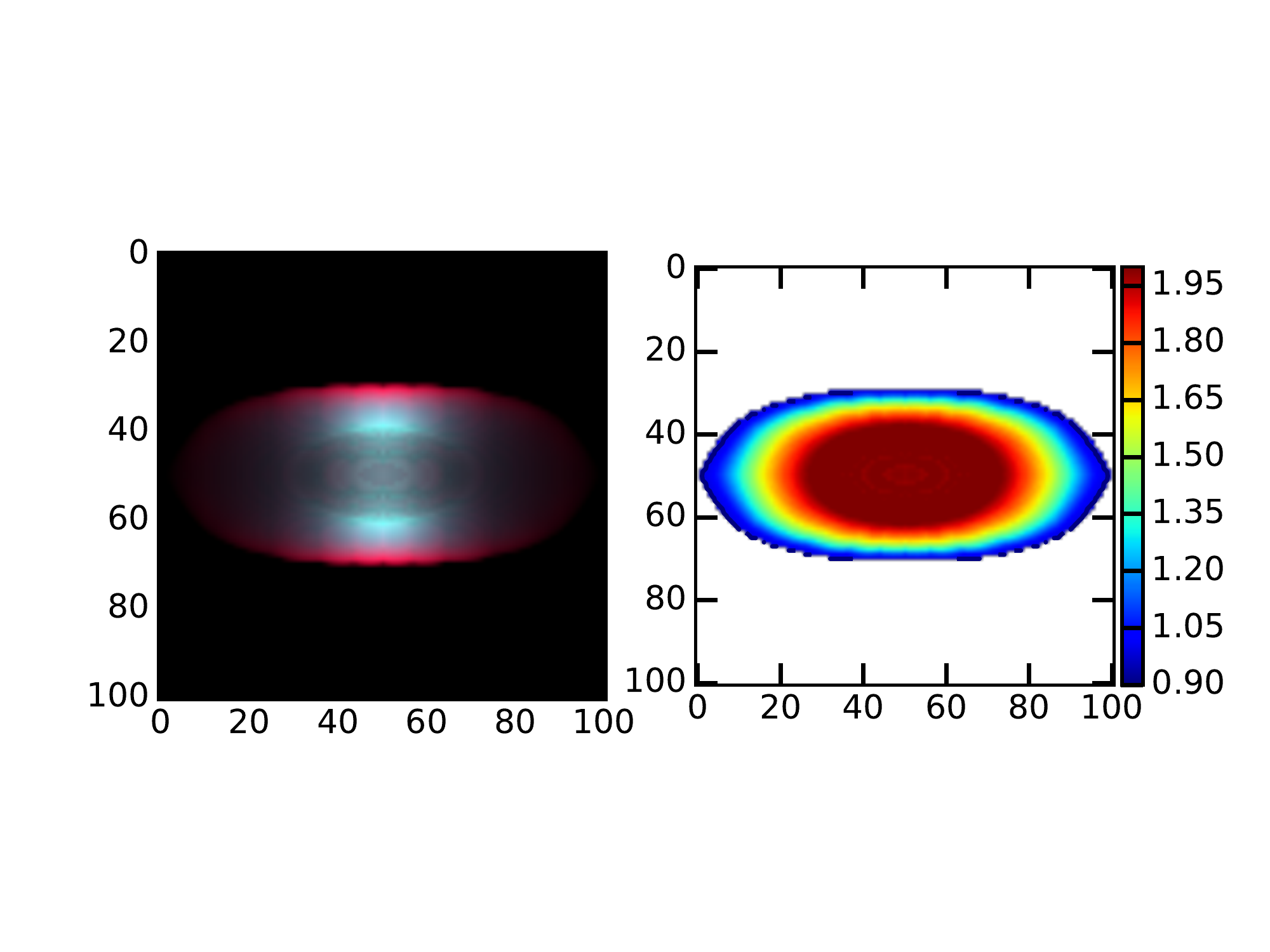} 
 \caption{Left: RGB image of the 3D toy model (Red: [\nii], Green: [\oiii], Blue: \hi). Right: ICF: the color bar is following  (1+N$^{++}$/N$^+$)/(1+O$^{++}$/O$^+$).}
   \label{fig:RGBICF}
\end{center}
\end{figure*}

3D models, even simple spherical ones, are useful if one wants to reproduce the observations obtained through slits that only partly overlap the nebula. This is especially true when one needs to use ICF to determine abundances, as the classical values have been determined for full nebulae. As an illustration, in Figure~\ref{fig:RGBICF} I show a simple 3D ellipsoidal model of a nebula ionized by a 50~kK star, with a mean ionization parameter of log(U)$=-2.5$. The left panel shows the main location of the low ionization (N$^+$ in red) and the high ionization (O$^{++}$ in green) regions. We can compute for each spaxel the ICF to apply to N$^+$/H$^+$ to obtain the N/H abundance, and compare it to the classical ICF (N/N$^+$ = O/O$^+$). This is shown in the right panel, where the color bar depicts any departure from 1.0 for this ratio. It is obvious that the ICF to be applied to N$^+$/H$^+$ is strongly dependent on the position of the aperture (the mean value of the departure (1+N$^{++}$/N$^{+}$)/(1+O$^{++}$/O$^{+}$) over the whole nebula is 1.26).

\subsection{Nebular kinematic}

Once a 3D model is obtained, one can add a velocity field and compute the emission line profiles as seen through any aperture. The effect of the geometry on the determination of the velocity fields from line profiles have been studied by \citet{2008Morisset_rmxa44}.
By fitting imaging with 3D models and high resolution profiles with expansion velocities \citet{2016Gesicki_aap585} determine dynamical properties of 8 PNe.
One must notice the long standing effort of the Potsdam group to develop hydrodynamical photoionization models: emission line profiles are used by \citet{2014Schonberner_Astr335} to constrain 1D models of the expansion properties and the internal kinematics of PNe.

\subsection{Topologically equivalent models}

\begin{figure*}[ht]
\begin{center}
 \includegraphics[width=4.3in]{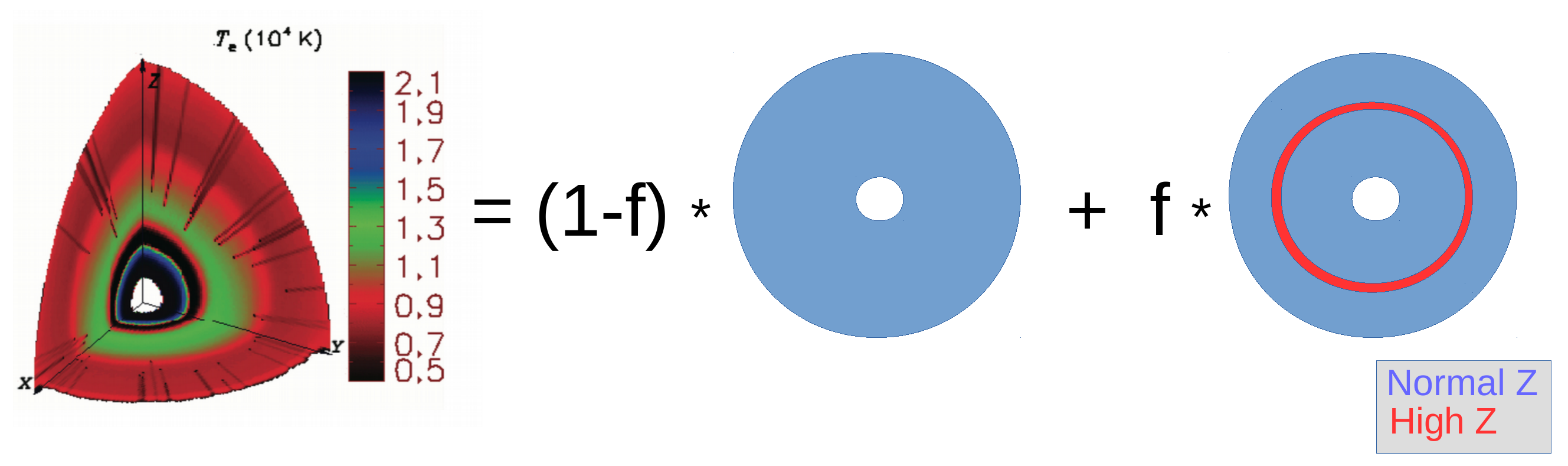} 
 \caption{Illustration of topologically equivalent models. Left image from \citet{2011Yuan_mnra411}.}
   \label{fig:topo}
\end{center}
\end{figure*}

One may wonder if a very detailed 3D model of a PN is actually needed if the nebula is not fully spherically symmetric. One can mentally reorganize the density distribution to a much simpler morphology that would almost have the same properties. As an illustration, in Figure~\ref{fig:topo}, I show that the 3D model of NGC 6153 by \citet{2011Yuan_mnra411} [made of high metallicity inclusions embedded into a normal metallicity diffuse component] could have been equivalently obtained with two runs of a 1D code: the first one with only the normal background component and the second one including a thin hydrogen-poor shell. The result of the combined model is obtained as the weighted sum of the line intensities of each individual model (with $f$ being the covering factor of the metallic clumps). This is exactly what has been done by \citet{2002Pequignot_12} in the first study of metallic inclusions in NGC 6153.
One can elaborate a more complex model by changing the radius of the metallic shell and summing up more that two models. The extrapolation of this method combined with a reconstruction of a given morphology leads to what we term pseudo-3D models \citep{2005Morisset_mnra360}. The limitations of the pseudo-3D models are reached when non-radial radiation dominates an important process. This is the case in only two situations: shadows behind optically thick clouds\footnote{There is no shadow behind the optically thin metallic clouds used by \citet{2011Yuan_mnra411}.} \citep{2004Morisset_313} and a region ionized by multiple and well separated sources (which is not relevant in the PNe case). Otherwise, pseudo-3D models are totally legitimate to model complex structures.
The simplicity of this approach and the gain in CPU time execution (from hours/days for the full 3D case to minutes for the equivalent or pseudo-3D model) will give the user alternative options to explore more of the space parameter and converge to a solution in a faster and more economic way as well as address the question of multiple solutions.

\subsection{Hidden inhomogeneities}

\begin{figure*}[ht]
\begin{center}
 \includegraphics[width=5.4in]{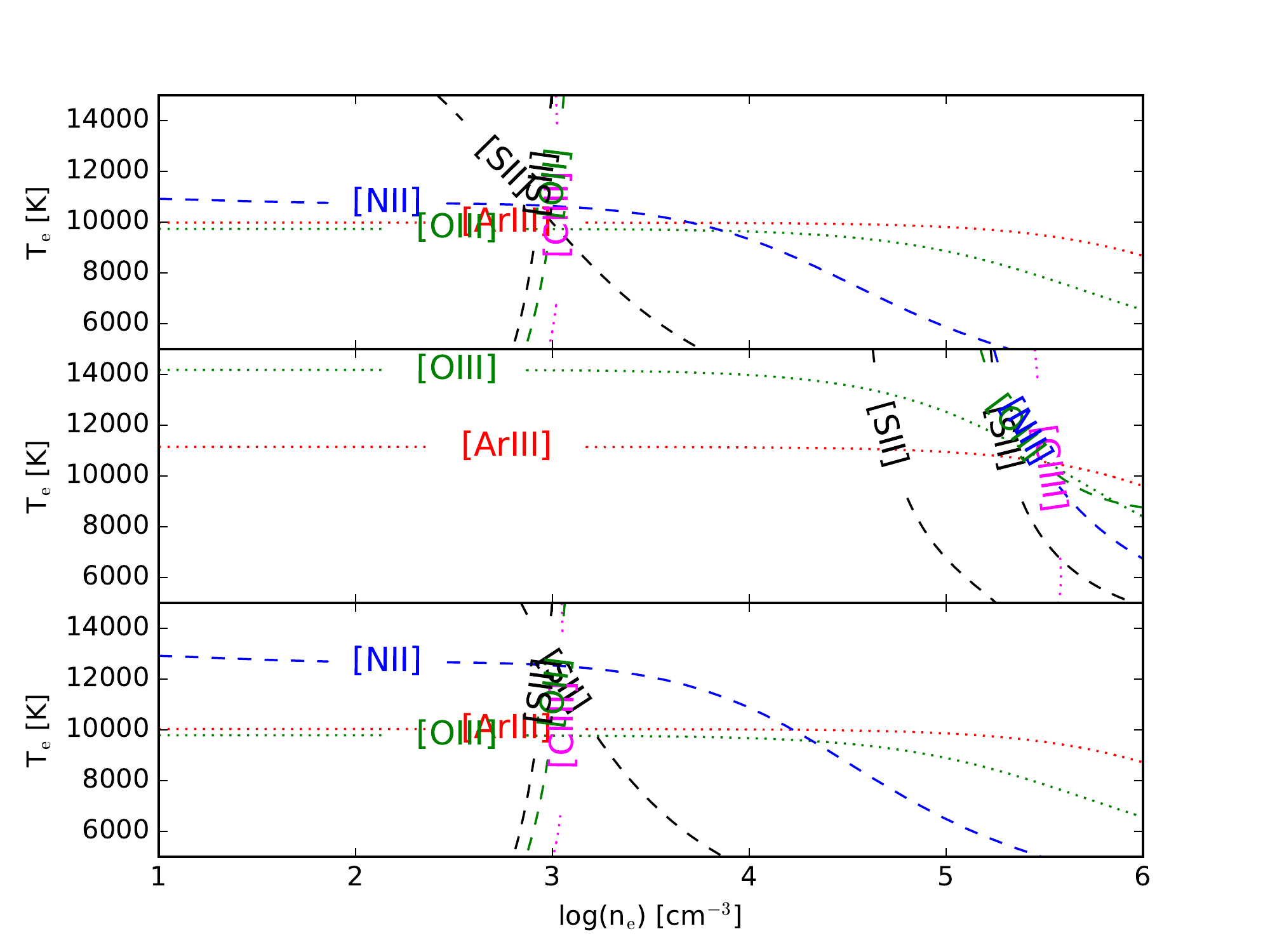} 
 \caption{Example of a 2-component model. From upper to lower panel: diagnostic diagrams for the low density medium (n$_{\rm H}$=10$^3$cm$^{-3}$, log(U) = -1.4, T$_{\rm e}\sim 9,700$~K), high density medium (n$_{\rm H}$=10$^{5.5}$cm$^{-3}$, log(U) = -3.2, T$_{\rm e}\sim 10,500$~K) and the composite model (1\% of the volume occupied by the high density medium).}
   \label{fig:2comps}
\end{center}
\end{figure*}

The fact that on the same line of sight two or more different components of gas are mixed leads to wrong results if trying to interpret the observations as coming from a single component. This is illustrated in Figure~\ref{fig:2comps}, where Te-Ne-diagnostic diagrams are shown for two models representing regions ionized by the same star but differing in density (n$_{\rm H}$=10$^3$ and 10$^{5.5}$ cm$^{-3}$ in the upper and middle panel respectively). The classical diagnostics line ratios are showing this difference in density, as well as the rather small difference in electron temperature (9,700 and 10,500~K resp.). But if one combines these two models into a single observation and repeat the same diagram (lower panel), it becomes impossible to detect the high density clumps and furthermore an apparent high temperature (13,000~K) is inferred from the low ionization [NII] emission lines, while the nebula is almost isothermal. This will in turn lead to wrong determinations of ionic and elemental abundances.

\section{Conclusions}

Photoionization codes are simply physical codes. They don't know about astrophysics. A set of input parameters that does not correspond to any realistic astrophysical situation will still lead a result in most situations, so be careful! It is the responsibility of the user to check the validity of the input parameters.

It is also the responsibility of the user to take care of the aperture effects and not compare a full model with observations obtained with a slit which is small compared to the size of the nebula. This is especially true when dealing with IFUs observations.

Topologically equivalent models can save a lot of CPU time. Finally, carefully exploring the space parameter is more important than fitting second-order parameters.
\newpage
\acknowledgments{}
Participation to this conference has been possible due to grant PAPIIT (DGAPA-UNAM) 107215. 
Many thanks to Daniel P\'equignot who taught me 20 years ago the basics (and a lot more) of photoionization models, especially the topologically equivalent models. Many thanks to Gra{\.z}yna Stasi\'nska who during the last 10 years helped me a lot in understanding in even greater details. Many thanks also to Gra{\.z}yna Stasi\'nska, Jorge Garc{\'{\i}}a-Rojas and Luc Binette for reviewing this paper.


\end{document}